\documentclass[aps,prl,floatfix,twocolumn,showpacs]{revtex4}%
\usepackage{amssymb}
\usepackage{amsmath}
\usepackage{graphicx}
\usepackage{amsfonts}%
\providecommand{\U}[1]{\protect\rule{.1in}{.1in}}
\begin{document}
\title{{A spin quantum bit with ferromagnetic contacts for circuit QED}}
\author{Audrey Cottet\thanks{To whom correspondence should be addressed :
cottet@lpa.ens.fr} and Takis Kontos}
\affiliation{Laboratoire Pierre Aigrain, Ecole Normale Sup\'{e}rieure, CNRS (UMR 8551),
Universit\'{e} P. et M. Curie, Universit\'{e} D. Diderot, 24 rue Lhomond,
75231 Paris Cedex 05, France}
\date{\today}
\pacs{03.67.Lx,73.23.-b,32.80.-t}

\begin{abstract}
We theoretically propose a scheme for a spin quantum bit based on a double
quantum dot contacted to ferromagnetic elements. Interface exchange effects
enable an all electric manipulation of the spin and a switchable strong
coupling to a superconducting coplanar waveguide cavity. Our setup does not
rely on any specific band structure and can in principle be realized with many
different types of nanoconductors. This allows to envision on-chip single spin
manipulation and read-out using cavity QED techniques.

\end{abstract}
\maketitle

Achieving a strong coupling between a single atom and a single photon trapped
in a cavity has been instrumental for studying the interaction of light and
matter at the most elementary level\cite{QED}. Recently, what was originally
belonging to the realm of quantum optics\cite{Raimond,T. Yoshie,J. P.
Reithmaier} has been brought into superconducting chips: two level
superconducting circuits (artificial atoms) have been embedded into
superconducting coplanar waveguide cavities, in the context of circuit Quantum
ElectroDynamics\cite{Wallraff}. Extending further this ``transfer'' to the
field of nano-fabricated circuits could offer a very interesting platform for
the study of entanglement and decoherence in many body systems.

\begin{figure}[ptb]
\includegraphics[width=1.0\linewidth]{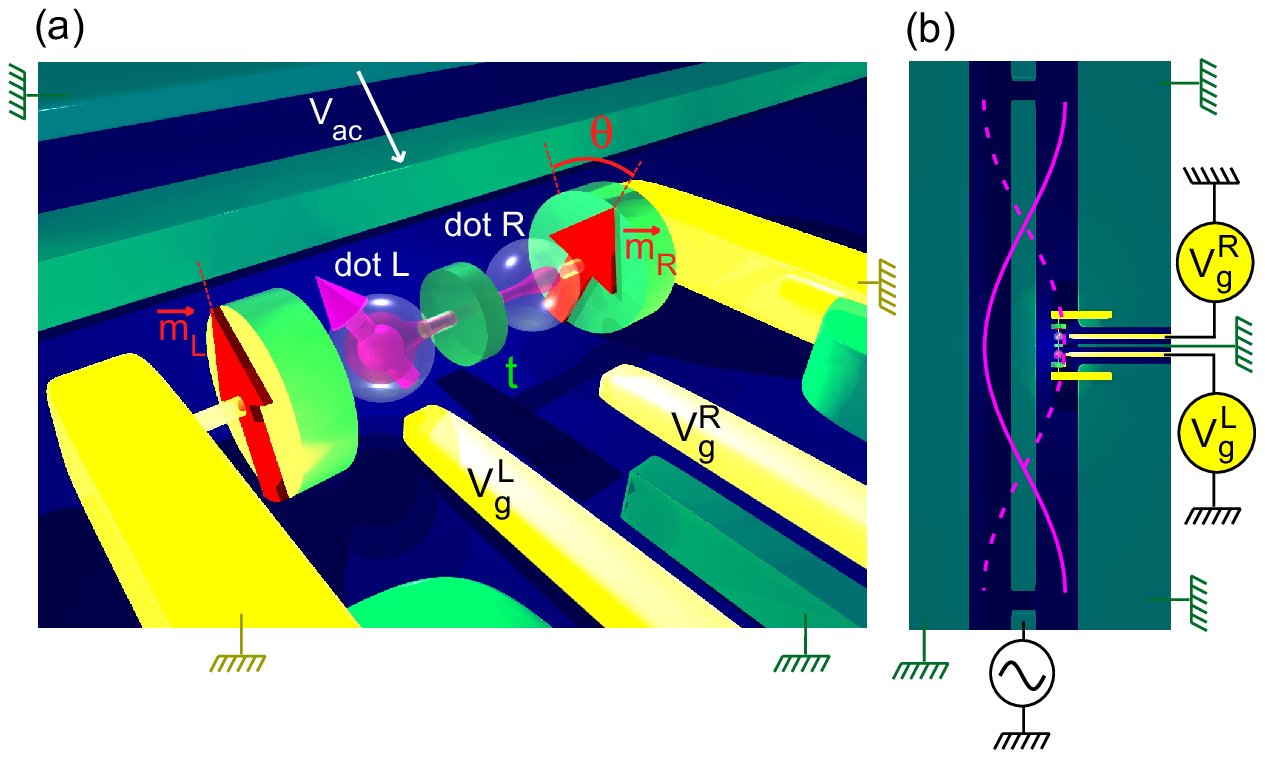}\caption{a.The two
quantum dots L and R (transparent spheres) are connected together through a
tunnel junction with hoping constant $t$ (small green disk). Each dot is
connected to a grounded normal reservoir (yellow rectangle) through a
ferromagnetic insulator ($FI$) (large green disks). The magnetizations
$\vec{m}_{L}$ and $\vec{m}_{R}$ of the two $FIs$ (red arrows) form an angle
$\theta$. The double dot circuit is placed between the center and ground
conductors of a superconducting coplanar waveguide (green tapes), which
conveys a voltage $V_{ac}$. The dot $L(R)$ is capacitively coupled to a gate
with voltage $V_{g}^{L(R)}$. We tune $V_{g}^{L}$ and $V_{g}^{R}$ such that
there is a single active electron on the double dot system, mainly localized
on dot L (magenta cloud). b. Top view of the same device on a wider scale. The
double dot is placed at an antinode of the cavity voltage standing wave. For
clarity we have not respected the relative scales.}%
\label{device}%
\end{figure}

The electronic spin of nanoconductors raises a strong interest in the fields
of quantum information and quantum coherence, because it is generally expected
to be more robust to decoherence than the orbital degrees of
freedom\cite{Hanson}. Coupling the spin of a nanoconductor to cavity photons
seems particularly attractive\cite{Childress}. Nanofabricated circuits are
naturally compatible with on-chip superconducting waveguide architectures.
However, two major difficulties have to be overcome. First, a strong coupling
$g$ between a spin and an electric field is intrinsically difficult to
achieve. Second, in architectures integrating several spins, the selective
manipulation of each spin is a challenge. The electrical schemes available for
local spin manipulation \cite{Tokura,Nowack,Trif1} and spin/photon
coupling\cite{Trif2,Burkard} rely on the use of a strong external
static\ magnetic field, which is not naturally compatible with strong photon
confinement in superconducting cavities\cite{Frunzio}. More generally, a
scheme allowing one to switch off $g$ efficiently is lacking, for
nanoconductor based qubits as well as superconducting qubits\cite{Girvin}. An
alternative way to spin-polarize the spectrum of a nanoconductor is to use
ferromagnetic insulator ($FI$) contacts, which have a very strong internal
Zeeman field. The hybridization of the nanoconductor orbitals with the first
atomic layers of the $FIs$ leads to an effective Zeeman field inside the
nanoconductor itself. This effect is quite universal since it can arise in
very different systems, such as carbon nanotubes \cite{Cottet2006,S. Sahoo,
Hauptman:08}, InAs quantum dots \cite{InAs} or thin superconducting
layers\cite{Tedrow,Cottet2009}. In this letter, we propose theoretically a
scheme for a spin qubit based on a double quantum dot contacted to
ferromagnetic insulators. Interface exchange effects enable an all electric
manipulation of the spin and a switchable strong coupling to a superconducting
cavity. First, our scheme is compatible with superconducting architectures
since it can be realized using $FI$ layer designs which minimize stray fields.
Second, it is generic in the sense that it does not rely on intrinsic
nanoconductor properties such as spin orbit coupling\cite{Trif2} or hyperfine
interaction\cite{Burkard}. Third, it is relatively robust since the
contact-induced effective fields can be adjusted in numerous ways, from the
fabrication to the operation of the quantum dots. This allows to envision
single spin manipulation and read-out using cavity QED techniques.

\begin{figure}[ptb]
\includegraphics[width=0.9\linewidth]{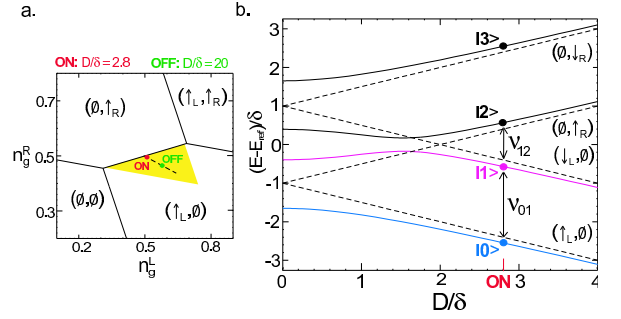}\caption{a. Stability
diagram indicating the double dot most stable state for $t=0$, versus the
reduced gate voltages $n_{g}^{L(R)}=C_{g}^{L(R)}V_{g}^{L(R)}/e$. We have used
the capacitance values indicated in the EPAPS, and $V_{ac}=0$. The circuit can
be operated along the dashed line $n_{g}^{R}=0.5+[(C_{\Sigma}^{R}%
+C_{m})/(C_{\Sigma}^{L}+C_{m})](0.5-n_{g}^{L})$ to maximize the energy
separation between the singly occupied levels $\left\vert \emptyset
,\uparrow_{R}\right\rangle $,$\left\vert \emptyset,\downarrow_{R}\right\rangle
$,$\left\vert \uparrow_{L},\emptyset\right\rangle $ and $\left\vert
\downarrow_{L},\emptyset\right\rangle $ and other energy levels corresponding
to a different double dot occupation. This yields a detuning $D=e^{2}%
(2n_{g}^{L}-1)/(C_{\Sigma}^{L}+C_{m})$ between the states $\left\vert
\uparrow_{L},0\right\rangle $ and $\left\vert 0,\uparrow_{R}\right\rangle $.
b. Eigenenergies of the double dot versus $D$ for $\delta=16\mu eV$,
$\theta=\pi/6$, $t=0$ (dashed lines), and $t=2\delta/3$ (solid lines). When
$t=0$, the levels $\left\vert \downarrow_{L},0\right\rangle $ and $\left\vert
0,\uparrow_{R}\right\rangle $ cross at $D=2\delta$. This crossing is replaced
by an anticrossing for $t\neq0$ and $\theta\neq0[2\pi]$. At $t=2\delta/3$,
$D=2.8\delta$ and $\theta=\pi/6$ (ON point), we obtain transition frequencies
$\nu_{01}=7.68~\mathrm{GHz}$ and $\nu_{12}=4.41~\mathrm{GHz}$.}%
\label{spectro2}%
\end{figure}

We first consider two generic quantum dots $L$ and $R$ coupled through a
tunnel barrier with a hopping constant $t$ (see Fig. \ref{device}). Each dot
is connected to a grounded normal metal reservoir through a $FI$ layer, and
capacitively coupled to a DC gate with voltage $V_{g}^{L(R)}$, which allows to
tune the chemical potential of the dot. The double dot system is placed
between the center and ground conductors of a superconducting coplanar
waveguide cavity, which conveys a voltage $V_{ac}$. We take into account one
orbital level for each quantum dot. The dots have strong self and mutual
charging energies, so that, by tuning properly $V_{g}^{L}$ and $V_{g}^{R}$, we
reach a regime where the total number of electrons on the double dot system is
constantly equal to 1 (see Fig. \ref{spectro2}). The tunnel rates through the
$FIs$ are sufficiently small so that cotunneling processes between the dot and
the normal reservoirs can be neglected. Our key ingredient is the
contact-induced effective spin splittings $2\delta_{L(R)}$ in dots $L(R)$. The
effective fields in dots $L$ and $R$ are colinear to the magnetizations
$\vec{m}_{L}$ and $\vec{m}_{R}$ of the left and right $FIs$, which form an
angle $\theta$. For later use, we note $\left\vert e_{L},e_{R}\right\rangle $
a double dot state with the dot $L(R)$ in state $e_{L(R)}$. Here, $e_{L(R)}$
can be the empty state $\emptyset$ or a singly occupied state with a given
spin (for instance $\uparrow_{L(R)}$ and $\downarrow_{L(R)}$ refer to spin
states parallel and antiparallel to $\overrightarrow{m}_{L(R)}$ respectively).
With the above assumptions, the simplified double dot hamiltonian writes%

\begin{align}
\hat{H}  &  =-\frac{D}{2}\hat{\tau}_{3}\hat{\sigma}_{0}+t\hat{\tau}_{1}%
\hat{\sigma}_{0}-\delta_{L}\hat{\sigma}_{3}\frac{\hat{\tau}_{0}+\hat{\tau}%
_{3}}{2}\nonumber\\
&  -\delta_{R}(\cos(\theta)\hat{\sigma}_{3}+\sin(\theta)\hat{\sigma}_{1}%
)\frac{\hat{\tau}_{0}-\hat{\tau}_{3}}{2} \label{H}%
\end{align}
Here, $\hat{\tau}_{i}$ and $\hat{\sigma}_{i}$, with $i\in\{0,1,2,3\}$, refer
to the identity and Pauli matrices in the left/right orbital subspace and the
$(\uparrow_{L},\downarrow_{L})$ spin subspace respectively. We note $D$ the
energy shift between the states $\left\vert \uparrow_{L},\emptyset
\right\rangle $ and $\left\vert \emptyset,\uparrow_{R}\right\rangle $ for
$t=0$. This quantity can be tuned with the dots' gate voltages\cite{EPAPS}.
For simplicity, we assume $\delta_{L}=\delta_{R}=\delta$. We define an "ON"
working point $D=D_{ON}=2.8\delta$, with $\theta=\pi/6$, $t=2\delta/3$ and
$\delta=16\mu\mathrm{eV}\simeq3.9~\mathrm{GHz}$. At the ON point, we note
$\left\vert \mathbf{j}\right\rangle $, with $\mathbf{j}\in\{\mathbf{1}%
,\mathbf{2},\mathbf{3},\mathbf{4}\}$, the eigenstates of $\hat{H}$, with
energies $E_{\mathbf{j}}$ growing with the index $\mathbf{j}$. The two lowest
eigenstates $\left\vert \mathbf{0}\right\rangle $ and $\left\vert
\mathbf{1}\right\rangle $ both have an overlap to the right dot of
only$\ \sim5\%$, and they correspond to $\uparrow_{L}$ and $\downarrow_{L}$
spin states to a good approximation since we obtain

\begin{figure}[ptb]
\includegraphics[width=0.5\linewidth,angle=-90]{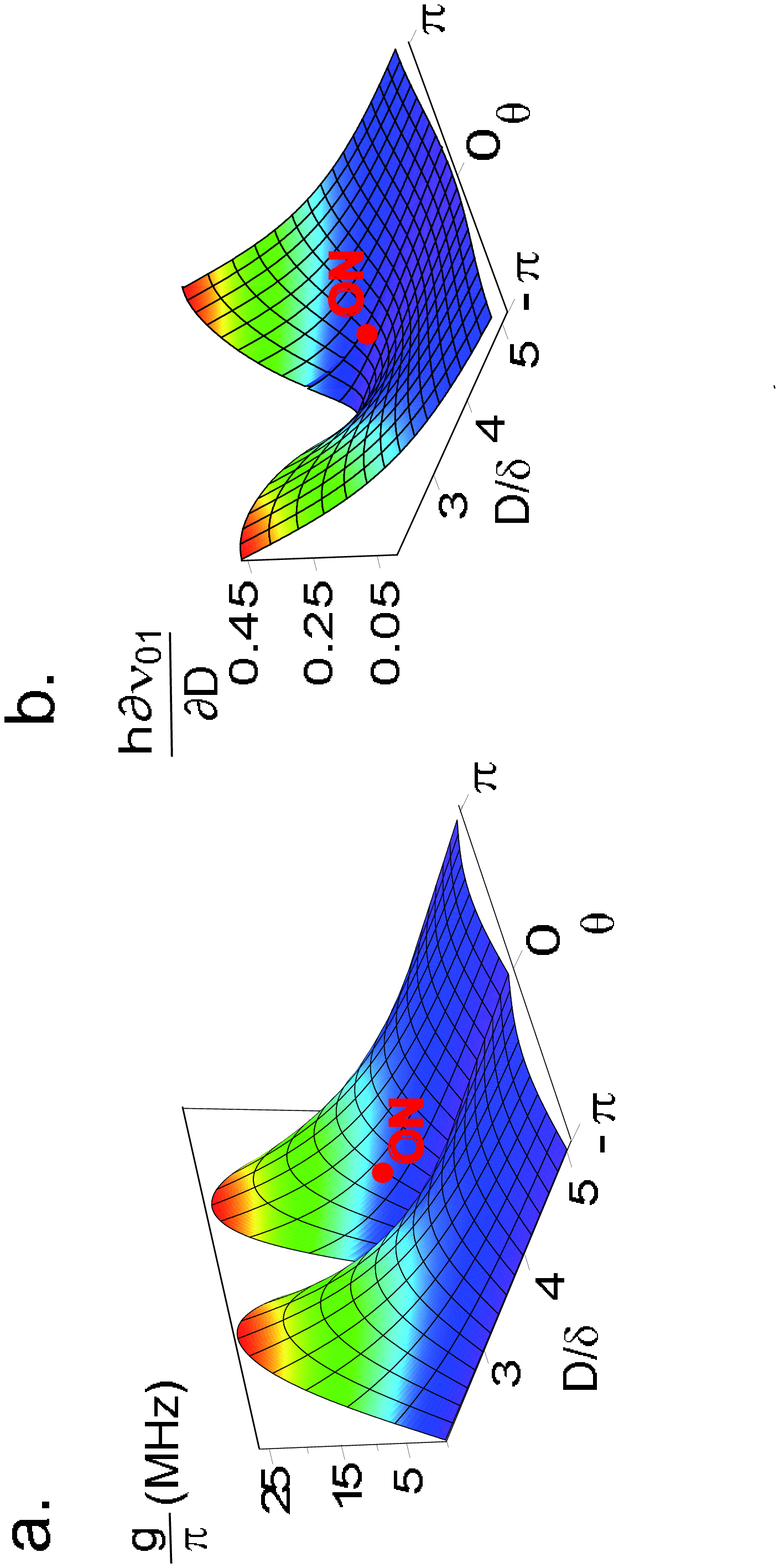}\caption{ a.
Spin-electric field coupling $g$ of the spin qubit, in terms of vacuum Rabi
frequency $g/\pi$. This quantity decreases with $D$, vanishes for
$\theta=0[\pi]$, and reaches a maximum for values $\theta=\pm\theta_{\max}(D)$
close to $\pm\pi/2$. At the ON point $(D=2.8\delta$, $\theta=\pi/6)$, we
obtain $g/\pi=5.6~\mathrm{MHz}$. At the OFF point ($D=20\delta$, $\theta
=\pi/6)$, which is indicated in Fig.\ref{spectro2}.a, we obtain
$g=13~\mathrm{kHz}$. b. Derivative of the transition frequency $\nu_{01}$ with
respect to $D$. This quantity mainly determines the sensitivity of the qubit
to charge noise mediated by fluctuations of $D$.}%
\label{g}%
\end{figure}

$\left\vert \left\langle \emptyset,\downarrow_{L}\right\vert \left.
\mathbf{0}\right\rangle \right\vert ^{2}=0.05\%$, $\left\vert \left\langle
\downarrow_{L},\emptyset\right\vert \left.  \mathbf{0}\right\rangle
\right\vert ^{2}=0.005\%$, $\left\vert \left\langle \emptyset,\uparrow
_{L}\right\vert \left.  \mathbf{1}\right\rangle \right\vert ^{2}=0.8\%$ and
$\left\vert \left\langle \uparrow_{L},\emptyset\right\vert \left.
\mathbf{1}\right\rangle \right\vert ^{2}=0.1\%$. We obtain a transition
frequency $\nu_{01}=(E_{1}-E_{0})/h\simeq1.98~\delta=7.68~\mathrm{GHz}$
accessible with current microwave technology. Importantly, the transition
frequency $\nu_{12}=(E_{2}-E_{1})/h=4.41~\mathrm{GHz}$\textrm{\ }between the
states $\left\vert \mathbf{1}\right\rangle $ and $\left\vert \mathbf{2}%
\right\rangle $ is significantly different from $\nu_{01}$. The double dot
circuit is thus sufficiently anharmonic to be operated as an effective two
level system $\{\left\vert \mathbf{0}\right\rangle ,\left\vert \mathbf{1}%
\right\rangle \}$, called "spin qubit" in the following. When $D$ is changed
to $D=D_{ON}+\delta D$ and $\theta\neq0[\pi]$, one obtains a coupling element
$\left\langle \mathbf{0}\right\vert \hat{H}\left\vert \mathbf{1}\right\rangle
=\mathcal{C}\delta D$ because the direction of the overall field felt by the
electron slightly changes. Importantly, when the capacitances connected to
dots $L$ and $R$ are asymmetric, one has $\partial D/\partial V_{ac}\neq0$
\cite{EPAPS}. This property combined with $\mathcal{C}\neq0$ allows a
transverse coupling of the spin qubit to the electric field conveyed by the
waveguide. In these conditions, Rabi oscillations between the states
$\left\vert \mathbf{0}\right\rangle $ and $\left\vert \mathbf{1}\right\rangle
$ can be obtained by applying an oscillating voltage with frequency $\nu_{01}$
to the waveguide center conductor. Using the rotating wave approximation and
the quantization rule $V_{ac}=V_{rms}(a+a^{\dag})$\cite{Blais}, with
$a,a^{\dag}$ the annihilation and creation operators for photons of the
waveguide cavity, the set of the spin qubit and cavity can be described with a
standard Jaynes-Cummings hamiltonian $\hat{H}_{eff}=-h\nu_{01}\hat{S}%
_{z}/2+\hbar\omega_{r}a^{\dag}a+\hbar g(a^{\dag}\hat{S}_{-}+a\hat{S}_{+})$
with $\hbar g=\mathcal{C}V_{rms}\partial D/\partial V_{ac}$, $\hat{S}_{\pm
}=(\hat{S}_{x}\pm i\hat{S}_{y})/2$ and $\hat{S}_{x}$, $\hat{S}_{y}$, $\hat
{S}_{z}$ the Pauli operators in the subspace $\{\left\vert \mathbf{0}%
\right\rangle ,\left\vert \mathbf{1}\right\rangle \}$. Using realistic values
for the circuit capacitances\cite{EPAPS} and $V_{rms}=2~\mu\mathrm{V}$, we
obtain a strong spin/electric field coupling at the ON point, i.e.
$g/\pi=5.6~\mathrm{MHz}$ in terms of vacuum Rabi frequency, a value comparable
to that obtained in Ref.$~$\onlinecite{Wallraff}. With such a coupling, the
readout of the qubit state can be performed by measuring the dispersive shift
of the cavity resonance frequency\cite{Blais}. Besides, a distant coupling
between two qubits can be realized by exchange of virtual cavity
photons\cite{Majer}. Importantly, our setup allows us to strongly decrease the
spin/cavity coupling by increasing $D$. For instance, we obtain $g/\pi
=13~\mathrm{kHz}$ at the OFF point $D=D_{OFF}=20\delta$ where $\nu_{01}%
\simeq2\delta=7.76~\mathrm{GHz.}$ Outside the qubit manipulation stage, the
qubit can be placed at the OFF point to minimize decoherence due to the
coupling to the cavity\cite{Houck}. This also allows us to switch off very
efficiently the coupling between two qubits via virtual cavity photons, with
almost no change of the qubit transition frequency. Then, the qubit has to be
placed at the ON point for the manipulation of its quantum state. This can be
performed by using appropriate gate voltage pulses. Note that $g$ also depends
on the angle $\theta$ as shown in Fig. \ref{g}.a. However, using a constant
$\theta$ imposed e.g. by the shape of the $FI$ layers should be sufficient in practice.

\begin{figure}[ptb]
\includegraphics[width=0.7\linewidth]{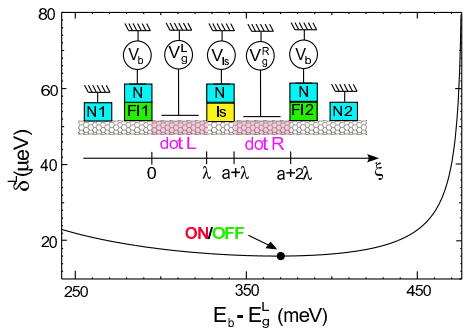}\caption{ Inset:
Implementation of our setup with a SWNT. The dots $L$ and $R$ are delimited by
the gates enclosing the insulating layer $Is$ and the ferromagnetic insulators
$FI1$ and $FI2$. The outer normal metal contacts $N1$ and $N2$ are used to
control the dots' occupation numbers. Main frame: Half effective Zeeman
splitting $\delta^{L}$ calculated for dot $L$ in the framework of
Eq.(\ref{Hnt}), with $\left\vert \uparrow_{L},\emptyset\right\rangle $ and
$\left\vert \downarrow_{L},\emptyset\right\rangle $ corresponding to the 29th
lowest pair of spin-dependent levels in the dot conduction band. This quantity
depends on the chemical potential shift $E_{b}-E_{g}^{L}$ between dot $L$ and
the neighbouring SWNT sections. We obtain a sweet spot $\partial\delta
^{L}/\partial E_{g}^{L}=0$ (black point) which can be used to reduce charge
noise-induced dephasing mediated by fluctuations of $\delta^{L}$. In this
Figure, we use $E_{b}=E_{Is}$, $R=1~\mathrm{nm}$, $\lambda=100~\mathrm{nm}$,
and $E_{ex}=3.7~\mathrm{meV}$. }%
\label{NT}%
\end{figure}

Reaching the strong coupling regime requires that the coherence times of the
spin and the cavity photons are longer than $1/g$. Since our setup does not
use any external magnetic field, it is compatible with using a superconducting
cavity with a good quality factor. Nevertheless, the coherence time of the
spin qubit can be limited by various mechanisms causing dephasing and
relaxation. These mechanisms depend on the structural properties of the
nanoconductors used. In principle, our scheme can be implemented with various
types of nanoconductors, such as e.g. silicon nanowires\cite{Shaji,Jonker} or
SWNTs. So far, the latter are the most advanced nanoconductors for realizing
our setup. Semiconducting SWNTs are particularly suitable for defining double
quantum dots, using top gates which allow one to tune the interdot coupling
$t$\cite{Mason}. Contact-induced Zeeman effective fields have already been
observed in SWNTs\cite{S. Sahoo, Hauptman:08}. Besides, multi-dot systems with
ferromagnetic leads and a local gate control have already been realized in the
open regime\cite{CPF}. We consider a SWNT above which several gates and
contacts are evaporated (inset of Fig.~\ref{NT}). The central electrostatic
gate enclosing an insulating layer $Is$ is used to define $t$. The two dots
are delimited by $Is$ and two other gates which enclose ferromagnetic
insulating layers $FI1$ and $FI2$. Here, the role the $FI$ layers of our
generic setup is played by the SWNT sections below $FI1$ and $FI2$. Following
the approach of Ref.~\onlinecite{Dani}, these SWNT sections are subject to an
effective Zeeman field $E_{ex}$. The electron confined in dot $L(R)$ feels
$E_{ex}$ through its evanescent tail, which leads to $\delta_{L(R)}\neq0$. We
describe the electronic bands of the SWNT with the Dirac-like hamiltonian
\begin{equation}
H_{W}=-i\hbar v_{F}(\frac{\hat{s}_{1}\hat{\gamma}_{3}}{R}\frac{\partial
}{\partial\varphi}+\hat{s}_{2}\frac{\partial}{\partial\xi})+\Delta
_{K-K^{\prime}}\hat{\gamma}_{1}-E_{pot}(\xi)\label{Hnt}%
\end{equation}
Here, $\hat{s}_{i}$ and $\hat{\gamma}_{i}$ with $i\in\{1,2,3\}$ are the Pauli
matrices on the A/B atoms and $K/K^{\prime}$ valley subspaces, $R$ is the
radius of the SWNT and $v_{F}=8~10^{5}m.s^{-1}$ is the SWNT Fermi velocity.
The term $\Delta_{K-K^{\prime}}\sim3~\mathrm{meV}$ describes a coupling
between the $K$ and $K^{\prime}$ bands, responsible for an orbital level
splitting currently observed in experiments\cite{Liang,Sapmaz}. We note $\xi$
and $\varphi$ the longitudinal and azimuthal SWNT coordinates. To obtain the
spectral properties of dots $L$ and $R$ separately ($t=0$), we use
$E_{pot}(\xi<0)=\sigma E_{ex}/2+E_{b}$, $E_{pot}(0<\xi<\lambda)=E_{g}^{L}$ and
$E_{pot}(\xi>\lambda)=E_{Is}$ for dot $L$, and $E_{pot}(\xi<a+\lambda)=E_{Is}%
$, $E_{pot}(a+\lambda<\xi<a+2\lambda)=E_{g}^{R}$, and $E_{pot}(\xi
>a+2\lambda)=\sigma E_{ex}/2+E_{b}$ for dot $R$. These potential profiles can
be adjusted with the various electrostatic gates shown in Fig.~\ref{NT}.
Remarkably, we find that $\delta^{L(R)}$ depends on $E_{ex}$, but also on the
size and the active orbitals of dots $L(R)$. Therefore, our scheme can work
for a wide range of $E_{ex}$ because $\delta^{L(R)}$ can be easily adjusted to
the value $\sim16~\mu\mathrm{eV}$ required in the generic model. Here, we use
$E_{ex}=3.7~\mathrm{meV}$, consistently with Ref.~\onlinecite{Dani},
$R=1~\mathrm{nm}$ and $\lambda=100~\mathrm{nm}$. We define the states
$\left\vert \uparrow_{L},\emptyset\right\rangle $ and $\left\vert
\downarrow_{L},\emptyset\right\rangle $ [$\left\vert \emptyset,\uparrow
_{R}\right\rangle $ and $\left\vert \emptyset,\downarrow_{R}\right\rangle $]
as the $29^{th}$ lowest pair of spin-dependent levels in the conduction band
of dot $L[R]$. Remarkably, for $E_{b}-E_{g}^{L}\sim1.05\Delta_{SWNT}$, with
$\Delta_{SWNT}=350~\mathrm{meV}$ the SWNT bandgap, we obtain a sweet spot
$\partial\delta^{L}/\partial E_{g}^{L}=0$, with $\delta^{L}=16~\mu\mathrm{eV}%
$. One has $E_{g}^{R}-E_{g}^{L}\ll E_{b}-E_{g}^{R(L)}$ at the ON/OFF points,
thus $\delta^{R}\sim\delta^{L}$. Finally, the interdot coupling $t$ can be
calculated from the overlap between the dots' orbitals. The value
$t=2\delta/3$ can be obtained by varying $a$ and $E_{Is}$, without affecting
significantly the sweet spot (not shown).

The spin coherence time is expected to be comparably long in SWNTS. Current
estimates based on the spin-orbit interaction in $^{13}$C-free SWNTs lead to
milliseconds\cite{Bulaev}. Since our setup provides a new way to control the
spin degree of freedom, it is a priori subject to specific decoherence
mechanisms which we evaluate below. First, low frequency charge noise picked
up by the dot gates risks to induce fluctuations of $D$ and thus $\nu_{01}$,
which will limit the spin dephasing time $T_{\varphi}^{D}$. Since $\partial
\nu_{01}/\partial D$ decreases with $D$, the choice of $D_{ON}$ results from a
compromise between having a low $T_{\varphi}^{D}$ and a large $g$. This
compromise also requires to use a relatively small value for $\theta$,
although $g$ is maximal for values of $\theta$\ close to $\pm\pi/2$ (see Fig.
\ref{g}.b). Using a semiclassical approach, we estimate $T_{\varphi}^{D}%
\sim2.9~\mathrm{\mu s}$ and $T_{\varphi}^{D}\simeq2~\mathrm{ms}$ at the ON and
OFF points, using the charge noise amplitude measured in non-suspended
SWNTs\cite{Herrmann}. Second, from Eq. (\ref{H}), $\nu_{01}$ corresponds to
$\sim2\delta^{L}$. Hence, charge noise can also dephase the qubit through
fluctuations of $\delta_{L}$. However, it is possible to minimize this effect
by working at the sweet spot $\partial\delta_{L}/\partial E_{g}^{L}=0$ in both
the ON and OFF states (see Fig.~\ref{NT}). This gives a dephasing time
$T_{\varphi}^{\delta_{L}}=15~\mathrm{ms}\gg T_{\varphi}^{D}$. Third, the
electron/phonon interaction conserves spin. However, since the states
$\left\vert \mathbf{0}\right\rangle $ and $\left\vert \mathbf{1}\right\rangle
$ have slight components in $\downarrow_{L}$ and $\uparrow_{L}$ respectively,
they can relax due to electron/phonon interaction. To evaluate this effect, we
take into account the acoustic mode most strongly coupled to electrons, i.e.
the stretching mode\cite{Ando}. We assume that the phonons in dots $L$ and $R$
are confined and decoupled due to the massive top gates. We obtain $T_{1}%
^{ON}=1.0~\mathrm{\mu s}$ and $T_{1}^{OFF}=0.21~\mathrm{s}$ at the ON and OFF
point respectively\cite{EPAPS}. Quite generally, using suspended
nanoconductors could reduce further decoherence, by decreasing the charge
noise picked up by the device, and also by reducing the phonon-induced $T_{1}$
thanks to a phononic Purcell-like effect. For instance, we estimate
$T_{1}^{ON}\simeq14~\mathrm{\mu s}$ and $T_{1}^{OFF}\simeq2.8~\mathrm{s}$ in
the SWNT setup, for phonons confined inside dots $L$ and $R$ with a quality
factor $Q_{ph}=20$. In principle, one can also expect relaxation due to the
coupling to the cavity, similar to what observed with superconducting
qubits\cite{Houck}. However, at the OFF point, we expect the cavity to be a
marginal source of decoherence since $g$ is almost switched off. At the ON
point, it is in principle possible to reduce the cavity-induced relaxation by
improving the cavity design\cite{Wang}. Hence, we have focused on decoherence
sources more specific to our qubit design. In summary, for a non-suspended
SWNT setup, we finally obtain a promising total decoherence time $T_{2}%
^{ON}=1.2~\mathrm{\mu s}\gg1/g=177~\mathrm{ns}$ at the ON point. Remarkably,
at the OFF point, we obtain $T_{2}^{OFF}\simeq2~\mathrm{ms}$ while $g$ is
reduced by a factor of $450$ compared to the ON point. This suggests that our
spin qubit could also be used as a quantum register.

\begin{acknowledgments}
We gratefully acknowledge fruitfull discussions with J. M. Raimond, B.
Dou\c{c}ot and H. Jaffr\`{e}s. This work was financed by the CNano Ile de
France contract SPINMOL.
\end{acknowledgments}

\end{document}